\documentclass[modern]{aastex63}
\newcommand{\rateunit}{\textrm{cm}^3/\textrm{s}}
\newcommand{\invcm}{\textrm{cm}^{-1}}
\newcommand{\cation}{\textrm{C}_2^+}
\newcommand{\Catom}{\textrm{C}(^3P)}
\newcommand{\Cion}{\textrm{C}^+(^2P^o)}

\received{July 19, 2019}
\revised{August 12, 2019}
\accepted{September 9, 2019}
\submitjournal{ApJ}

\shorttitle{Radiative association of atomic and ionic carbon}
\shortauthors{Babb, Smyth, and McLaughlin.}
\begin{document}

\title{Radiative association of atomic and ionic carbon}

\correspondingauthor{James F. Babb}
\email{jbabb@cfa.harvard.edu, rsmyth41@qub.ac.uk, bmclaughlin899@btinternet.com}

\author[0000-0002-3883-9501]{James F. Babb}
\affil{Center for Astrophysics \textbar \ Harvard  Smithsonian, 
       60 Garden St., MS 14,
       Cambridge, MA 02138}

\author[0000-0002-4359-1408]{R. T. Smyth}
\affil{Center for Astrophysics \textbar \ Harvard  Smithsonian, 
       60 Garden St., MS 14,
       Cambridge, MA 02138}
\affil{Centre for Theoretical Atomic, Molecular, 
        and Optical Physics, School of Mathematics \& Physics \\ 
        Queen's University of Belfast, Belfast BT7 1NN, Northern Ireland, UK}

\author[0000-0002-5917-0763]{B. M. McLaughlin}
\affil{Center for Astrophysics \textbar \ Harvard  Smithsonian, 
       60 Garden St., MS 14,
       Cambridge, MA 02138}
\affil{Centre for Theoretical Atomic, Molecular, 
       and Optical Physics, School of Mathematics \& Physics \\ 
       Queen's University of Belfast, Belfast BT7 1NN, Northern Ireland, UK}

\begin{abstract}
We present calculated cross sections 
and rate coefficients for the formation 
of the dicarbon cation ($\cation$)
by the radiative association process in collisions of a $\Catom$ atom
and a $\Cion$ ion. 
Molecular 
structure calculations 
for a number of  
low-lying doublet and quartet states of $\cation$
are used to obtain
the potential energy surfaces and transition 
dipole moments coupling the states of interest,
substantially increasing the available molecular data for $\cation$.
Using a quantum-mechanical method, we explore a number of allowed transitions
and determine those
contributing to the radiative association process.
The calculations extend
the available data for this process down to the temperature of 100~K,
where the rate coefficient is found to be 
about $2\times 10^{-18}\,\rateunit$.
We provide analytical fits suitable for incorporation into
astrochemical reaction databases.
\end{abstract}

\section{Introduction}\label{sec:intro}
The dicarbon cation $\cation$ is an important molecule in astrochemistry,
as it is one of the species participating in hydrocarbon chemistry
in, for example, interstellar clouds~\citep{Solomon72} and photon-dominated regions~\citep{Guzman15}.
In  environments with hydrogen the dication is rapidly consumed, but
it is of interest to securely characterize $\cation$ formation mechanisms
as the subsequent production of larger molecules $\textrm{C}_2\textrm{H}^+$, $\textrm{C}_2\textrm{H}_2^+$, ...
will depend on the available $\cation$.

In the present paper we consider the formation of
the dicarbon cation ($\cation$) by the radiative association process in
collisions of a carbon atom and a carbon ion,
\begin{equation}
\label{RA}
\Catom + \Cion \rightarrow \cation + h\nu .
\end{equation}
The process (\ref{RA}) is a mechanism for dicarbon cation formation,
which is viable in astrochemical environments because of the applicability
of two-body kinetics and because of the abundance of carbon. (Another, more recent,
application is
to carbon plasma chemistry~\citep{hirooka2014}.)
We note---we will discuss the chemistry in more detail below---that in typical astrophysical applications
with relatively abundant hydrogen (in atomic and molecular form) the radiative reaction (\ref{RA}) will
generally be outpaced by the ionic reactions such as~\citep{Solomon72}
$\textrm{CH}   +\textrm{C}^+ \rightarrow \cation + \textrm{H}$~\citep{federman89,Guzman15}
or
$\textrm{CH}^+ +\textrm{C}  \rightarrow \cation + \textrm{H}$~\citep{Chabot13,Rampino16}
and (\ref{RA}) is likely to be a minor process.
Nevertheless, there is one previous calculation of (\ref{RA}) that was carried out
to a lowest temperature of 300~K and which is listed in astrochemical
reaction databases for applications
below this temperature.
To remove the uncertainty for astrochemical applications it is necessary
to calculate the rate coefficient at lower temperatures.
In the present paper we compute cross sections and rate coefficients
for (\ref{RA}) and provide a new  calculation of
the process valid down to 100~K. 
We find that the rate coefficient is larger for $100<T<300$~K than
the values assumed in astrochemical databases.

The theory of the  formation of diatomic species by radiative association
is generally established.
A recent review summarizes the theoretical methdologies and lists 73 diatomic species\footnote{We remark,
somewhat tangentially,
that there
are also listed 5 polyatomic species, not including, for example,
a recent calculation for the sodium ion and the hydrogen molecule~\citep{burdakova19}.} for which
calculations have been carried out~\citep{Nyman15}.
There is, however, only one previous calculation for $\cation$~\citep{andreazza97},
and that calculation does not extend below 300~K.
[A statistical (RRKM) theory applied to radiative association of $\textrm{C}^+$ and $\textrm{C}_n$
for $n=1,...,8$
yielded a value of zero for the rate coefficient of process (\ref{RA}) at 30~K,
but the theory was focused on the polyatomic case~\citep{Freed82}.]

In a molecular description of the collisions that
describe (\ref{RA}) there are numerous possible approach channels
and transitions leading to $\cation$
and the theoretical treatment of (\ref{RA}) becomes quite complicated,
as we will show.
The rate coefficients for (\ref{RA}) were
calculated by \citet{andreazza97} 
using a semi-classical description
of the collisions in which the  atom and ion 
approach in the  quartet $B\;{}^4\Sigma_u^-$ state, yielding
a rate coefficient of about $3\times 10^{-18}\,\rateunit$ at 300~K.
In this paper, we will show by calculating
the cross sections using a quantum-mechanical method with improved molecular
data for $\cation$ that the most significant
channels are those in which the colliding atom and ion approach in molecular doublet states.
We present calculations for a number of molecular states of $\cation$,
cross sections and rate coefficients for process (\ref{RA}),
discuss the present rate coefficients and the earlier determination,
and consider in more detail potential significance of (\ref{RA}).

\section{Molecular structure}\label{sec:struct}

Potential energies correlating to $\Catom$ and $\Cion$ 
were calculated by \citet{Petrongolo81,Rosmus86,bruna92,ballance01,Shi13}.
There are several experimental studies of band spectra including~\citet{maier88,boudjarane95,tarsitano04}
and calculations of transition dipole moment functions~\citep{Rosmus86}.
As far as we know, no new transition dipole moment 
calculations have appeared since \citet{Rosmus86} yet
for the present study we will require a larger set.
Therefore, we calculated a number of molecular potential energies
and corresponding transition dipole moments substantially increasing
the available data for $\cation$.

The potential energy curves (PECs) and transition dipole moments 
(TDMs) are calculated for a set of low-lying doublet and quartet electronic states
that enter into the radiative association calculations.
We treat the molecular states formed 
by the approach of $\Catom$ with 
$\Cion$ within an MRCI+Q approximation:
A state-averaged (SA) complete-active-space-self-consistent-field 
(SA-CASSCF) approach, followed by multi-reference 
configuration interaction (MRCI) calculations, together 
with the Davidson correction (MRCI+Q) \citep{Helgaker2000}, is used. 
The SA-CASSCF method is used as the 
reference wave function for the MRCI calculations. 
The basis sets used in the present work are the augmented correlation consistent 
polarized core valence quintuplet [aug-cc-pcV5Z (ACV5Z)] Gaussian basis sets,
which
were used in our recent work on the dicarbon molecule~\citep{Babb2019} and 
were found to give 
an excellent representation of the states as the molecule dissociated.

All the PEC and TDM calculations for 
$\cation$ were performed with the 
quantum chemistry program package 
\textsc{molpro} 2015.1 \citep{molpro2015}, 
running on parallel architectures.  
For molecules with degenerate symmetry, 
an Abelian subgroup is required to be used in 
\textsc{molpro}. Thus, for $\Cion$ with D$_{{\infty}h}$ symmetry, 
it will be substituted by D$_{2h}$ symmetry with the 
order of irreducible representations  being 
($A_g$, $B_{3g}$, $B_{2g}$, $B_{1g}$, $B_{1u}$, $B_{2u}$, $B_{3u}$,  $A_u$). 
When the symmetry is reduced from D$_{{\infty}h}$ to D$_{2h}$~\citep{herz50},
the correlating relationships are 
$\sigma_g \rightarrow a_g$, 
$\sigma_u \rightarrow a_u$, 
$\pi_g \rightarrow$ ($b_{2g}$, $b_{3g}$), 
$\pi_u \rightarrow$ ($b_{2u}$, $b_{3u}$), 
$\delta_g \rightarrow$ ($a_g$, $b_{1g}$),  and 
$\delta_u \rightarrow$ ($a_u$, $b_{1u}$). 

In order to take account of short-range interactions, 
we employed the non-relativistic SA-CASSCF/MRCI method 
available within the \textsc{molpro} 
\citep{molpro2012,molpro2015} quantum chemistry suite of codes.  
For the dicarbon cation,  molecular  
orbitals (MOs) are put into the active space, 
including (3$a_g$,  1$b_{3u}$, 1$b_{2u}$, 0$b_{1g}$, 
3$b_{1u}$, 1$b_{2g}$, 1$b_{3g}$, 0$a_u$),  symmetry MOs.
The molecular orbitals for the MRCI procedure were obtained using the 
SA-CASCSF method, for doublet and quartet spin symmetries.
The averaging processes was carried out on  the two lowest states 
of the symmetries: ($A_g$, $B_{3u}$, $B_{1g}$, $B_{1u}$)
and the lowest states of the symmetries; 
($B_{2u}$, $B_{3g}$, $B_{2g}$ and $A_u$).  
This gives an accurate representation of the doublet and quartet  
states.

There are 24 molecular electronic states formed from $\Catom$ and $\Cion$, namely,
$^{2,4}\Sigma_{g,u}^+$, $^{2,4}\Sigma_{g,u}^- (2)$, $^{2,4}\Pi_{g,u} (2)$, $^{2,4}\Delta_{g,u}$~\citep{Chiu73,Shi13},
however,
we do not consider
the  $2^2\Sigma_u^-$, $2^2\Sigma_g^-$, $1\,^4\Pi_u$,
$2\,^4\Sigma_u^-$, $1\,^4\Sigma_g^+$,
$2\,^4\Pi_u$, $1\,^4\Delta_g$ and
$2\,^4\Sigma_g^-$ states because they
are repulsive~\citep{Shi13}  and 
will not be important for process (\ref{RA}).

\begin{figure}[htp]
\plotone{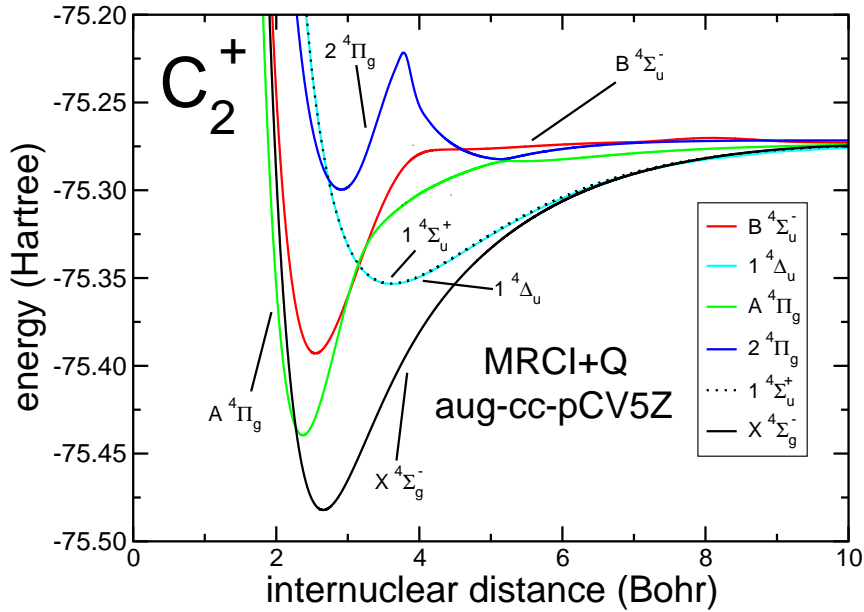}
\caption{Calculated  quartet molecular states of the $\cation$ cation
         as a function of internuclear distance $R$.
         We note that the 1$^4\Sigma_u^+$ and the 1 $^4\Delta_u$ states 
         are almost degenerate in energy.
\label{fig:pots-quartet}}
\end{figure}

\begin{figure}[htp]
\plotone{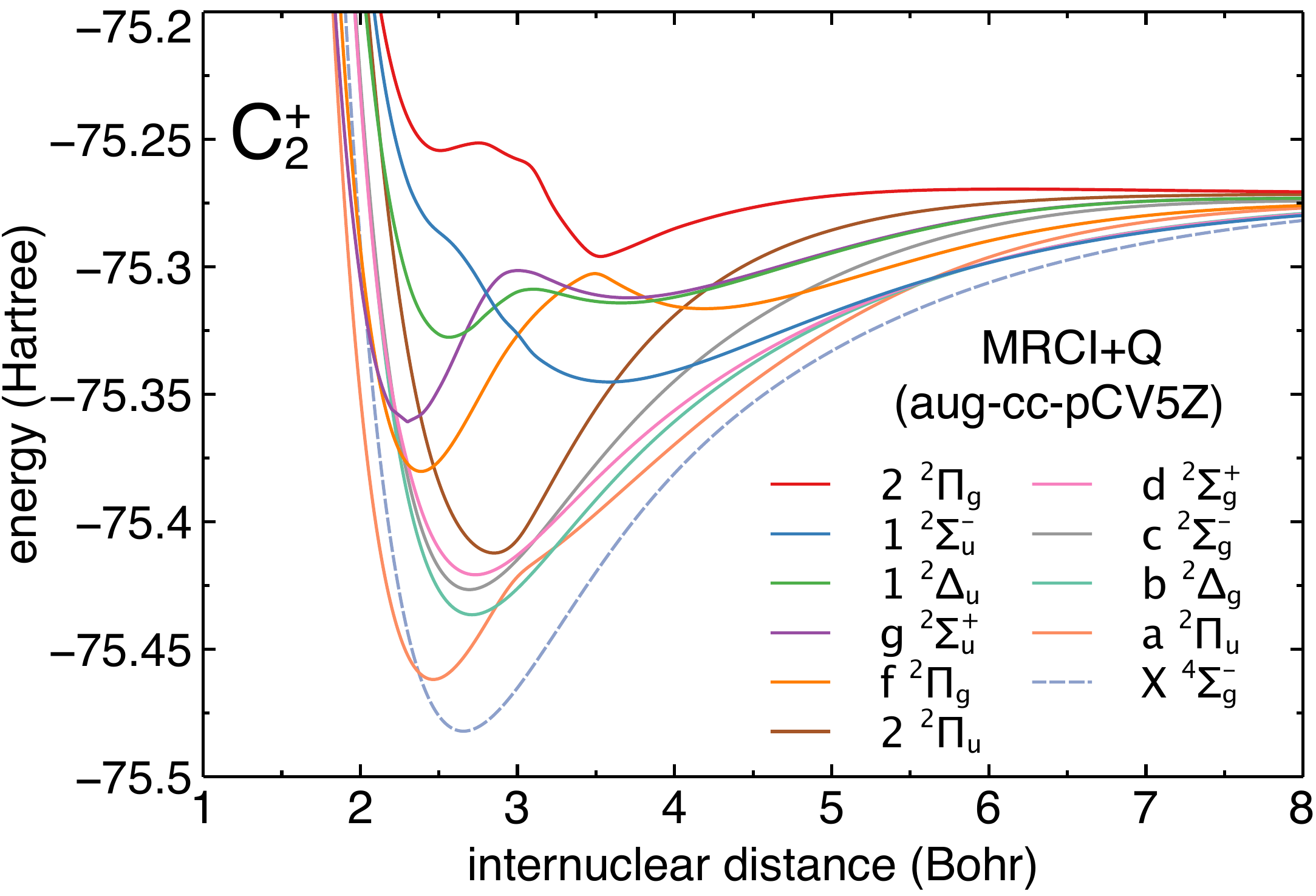}
\caption{Calculated doublet molecular states of $\cation$
         that were investigated in 
         our cross-section calculations, 
         as a function of internuclear distance $R$, with
         the X$^4\Sigma_g^-$ ground state indicated.
         In the index the states are listed  in order of decreasing
         energy at $R=2.6\,a_0$ starting from the left column:
         $2\,^2\Pi_g$, ..., $2\,^2\Pi_u$, $d\,^2\Sigma_g^+$, ..., $X\,^4\Sigma_g^-$.
\label{fig:pots-doublet}}
\end{figure}

Potential energy curves and transition dipole 
moments as functions of internuclear distance 
$R$ were calculated over the range $1.5\leq R \leq 20$~Bohr
for the sixteen states listed in Table~\ref{table:states}.
\begin{deluxetable}{c@{$\,$}lDD}[ht!]
    \centering
    \tablecaption{
    Theoretical $T_e$ values (units of $\invcm$) for the 16 states that were considered in the present work.
    In column 2 the values of the term energies $T_e$ are listed for
    the potential energies fit to Eq.~(\ref{eq:longrange}) relative to the minimum
    of the $\textrm{X}^4\Sigma_g^-$ potential energy and in column 3 the MRCI+Q/CV+DK+56
    values from \citep{Shi13} are given.
    \label{table:states}}
    \tablehead{
    \twocolhead{State} & 
    \twocolhead{Present\tablenotemark{a}} &
    \twocolhead{Other\tablenotemark{b}} 
        }
    \decimals
    \startdata
    X&$^4\Sigma_g^-$         & 0            & 0       \\
    a&$^2\Pi_u$              & 4417         & 4590.09   \\
    A&$^4\Pi_g$              & 9399         & 9597.8    \\
    b&$^2\Delta_g$           & 10027        & 9943.30    \\
    c&$^2\Sigma_g^-$         & 12295        & 12179.74   \\
    d&$^2\Sigma_g^+$         & 13281        & 13505.81    \\
    2&$^2\Pi_u$              & 15238        & 15366.74    \\
    B&$^4\Sigma_u^-$         & 19624        & 19767.42    \\
    f&$^2\Pi_g$              & 22332        & 22618.62     \\
    g&$^2\Sigma_u^+$         & 26433        & 26797.85   \\
    1&$^4\Delta_u$           & 28151        & 28425.26     \\
    1&$^4\Sigma_u^+$         & 28273        & 28859.38     \\
    1&$^2\Sigma_u^-$         & 30053        & 30174.91     \\
    1&$^2\Delta_u$           & 33942        & 34221.58   \\
    2&$^2\Pi_g$              & 40767        & 36448.37\tablenotemark{c}    \\
    2&$^4\Pi_g$              & 39987        & 40271.62     \\
    \hline
    \enddata
    \tablenotetext{a}{Present calculations.} 
    \tablenotetext{b}{From \protect\citet{Shi13} MRCI+Q/CV+DK+56.}
    \tablenotetext{c}{The value of the second well in the present calculation
    is $36262\,\invcm$, in agreement with the single well $36338.27\,\invcm$
    given by \citet{Shi13}. See text for discussion.}
    \end{deluxetable}
    %
    %
In Fig.~\ref{fig:pots-quartet} we present the calculated
quartet PECs.
In Fig.~\ref{fig:pots-doublet} we present
the calculated doublet PECs along with the $X\,^4\Sigma_g^-$ PEC
included for reference.

For  $R \geq 20$ Bohr, the PECs were fit to the long-range form
\begin{equation}\label{eq:longrange}
    V(R) = \frac{C_3}{R^3}-\frac{\alpha_d(\textrm{C})}{2R^4}\,,
\end{equation}
where $C_3/R^3$ is the electric
charge--atomic-quadrupole interaction potential energy~\citep{GenGie77} and $\alpha_d(\textrm{C})$ is the static
electric dipole polarizability of carbon~\citep{MilKel72,DasTha98} for each state. 
We utilized the $\alpha_d(\textrm{C})$  values calculated by \citet{DasTha98} with the finite field method 
and a coupled cluster CCSD(T) approach for atomic carbon
in the $M_L=0,\pm 1$ states. 
For example, we used $C_3=-0.775$ and $\alpha_d(\textrm{C})=12.4$ for the X$^4\Sigma_g^-$,
A$^4\Pi_g$, 1$^4\Delta_u$,
a$^2\Pi_u$, b$^2\Delta_g$,  c$^2\Sigma_g^-$, and f$\;^2\Pi_g$   states,
and we used $C_3=1.55$ and $\alpha_d(\textrm{C})=10.3$ for
the B$^4\Sigma_u^-$, 2$^4\Pi_g$, 2$^2\Pi_u$, and  2$^2\Pi_g$ states.
For $R \leq 1.5$, a short-range interaction potential 
of the form $\exp(AR)$ was fitted to 
the \textit{ab initio} potential curves.

The resulting $T_e$ values measured with respect to the minimum
of the $X^4\Sigma_g^-$ state are given in Table~\ref{table:states}.
We also list the values of $T_e$ from the
recent MRCI+Q/CV+DK+56 calculations of \citet{Shi13}.
[Comprehensive tabulations of calculations
by earlier workers (and some experimental data) were given by \citet{Shi13}.]
For $T_e$, there is generally very good agreement (within 300~$\invcm$)  between
our calculations and those of \citet{Shi13},
which also 
include relativistic (Douglas-Kroll effective Hamiltonian) effects and which
were obtained using extrapolation procedures to account for energy dependencies
on basis sets.
For example, we find $T_e=19624\,\invcm$ for the $B^4\Sigma_u^-$ state,
in very good agreement with the experimental value of $19652.2\pm0.4\,\invcm$~\citep{Celii1990},
while \citet{Shi13} find  
an MRCI+Q/CV value of $19718.80\,\invcm$ and an  
MRCI+Q/CV+DK+56  value\footnote{\protect\citet{Shi13} declare
``no other theoretical $T_e$ is closer to the measurements than'' (theirs). 
While our value is closer to experiment than theirs, without a 
detailed convergence study,  our close agreement 
with experiment may be coincidental. They also state ``no other theoretical 
$R_e$ result is superior to (theirs) in quality when compared 
with the measurements''. We obtain the same value of 
$R_e=2.55\,a_0$.} of $19767.42\,\invcm$.
For the  $c^2\Sigma_g^-$  state, we find $T_e=12295\,\invcm$, to be compared
to calculations by \citet{Shi13} listing
an MRCI+Q/CV value  of $12221.88\;\invcm$ and a CV+DK value of $12213.98\;\invcm$ (their
Table~3) and an  MRCI+Q/CV+DK+56 value of $12179.74\;\invcm$
(their Table~5).
Similarly, for the $a^2\Pi_u$ state, we find $T_e=4417\,\invcm$, while \citep{Shi13}
find  an MRCI+Q/CV+DK+56 value of $4590.09\,\invcm$.
In the case of the $2\,^2\Pi_g$ state, we find two wells with, respectively,
$T_e=40767$ and $36262\,\invcm$, whereas \citet{Shi13} list only a
single well of depth $T_e=36448.37\,\invcm$.
Close comparison of \citet[Fig.~1]{Shi13} and the present Fig.~\ref{fig:pots-doublet}
reveals that \citet{Shi13} plot (and list wells) corresponding 
to \textit{diabatic} potential curves for
the states of $^2\Pi_g$ symmetry, giving 
a single well for the $2\,^2\Pi_g$ PEC.
In the adiabatic approximation, which we utilize,
the $f\,^2\Pi_g$ and $2\,^2\Pi_g$ PECs do not cross~\cite[p. 295]{herz50}
and the $f\,^2\Pi_g$ PEC ``turns over''
as it approaches the $2\,^2\Pi_g$ state with increasing $R$ yielding a second well.
The avoided crossings in the present calculations between the $f\,^2\Pi_g$ and $2\,^2\Pi_g$ states and between 
the $2\,^2\Pi_u$ and $1\,^2\Pi_u$ states are in good agreement
with earlier work~\citep{Petrongolo81,ballance01}.
The accuracy of the present PECs is suitable for the present calculations
of the radiative association process.

There are numerous allowed electric dipole transitions within
the manifold of electronic states calculated here
and listed in Table~\ref{table:states}~\citep[p. 243]{herz50}.
The radiative association cross sections arise from
spontaneous transitions between the vibrational continuum
of the initial electronic state and
a bound vibrational state of the final electronic state
and depend roughly on the third
power of the electronic transition energies and the square of
the transition dipole moments.
Thus, we will investigate in further detail only 
the allowed transitions where the
electronic transition energies and transition dipole
moments are comparatively large.
In anticipation of the cross section calculations presented
in the next section, 
Table \ref{table:transitions} gives a list of the transitions 
investigated here for the radiative association process.  
\begin{deluxetable}{ccc}[ht!]
\centering
\tablecaption{Molecular transitions of the dicarbon cation 
              investigated in this work listed in order of 
              decreasing contribution to the total cross 
              section. }
\label{table:transitions}
\tablewidth{6in}
\tablehead{
\colhead{Initial state}  & to &
\colhead{Final state} 
                }
    \startdata
   $f\,^2\Pi_g$      &$\rightarrow$ &$a\,^2\Pi_u$   \\
   $2\,^2\Pi_g$      &$\rightarrow$  &$a\,^2\Pi_u$  \\
   $B\,^4\Sigma_u^-$ &$\rightarrow$  &$X\,^4\Sigma_g^-$  \\
   $b\,^2\Delta_g$   &$\rightarrow$  &$a\,^2\Pi_u$  \\
   $f\,^2\Pi_g$      &$\rightarrow$  &$2\,^2\Pi_u$   \\  
   $2\,^2\Pi_g$      &$\rightarrow$  &$2\,^2\Pi_u$   \\
   $1\,^2\Sigma_u^-$ &$\rightarrow$  &$f\,^2\Pi_g$      \\ 
   $c\,^2\Sigma_g^-$ &$\rightarrow$  &$a\,^2\Pi_u$   \\
   $B\,^4\Sigma_u^-$ &$\rightarrow$  &$A\,^4\Pi_g$   \\
   $2\,^2\Pi_u$      &$\rightarrow$  &$c\,^2\Sigma_g^-$  \\
   $2\,^4\Pi_g$      &$\rightarrow$  &$1\,^4\Sigma_u^+$  \\ 
   $2\,^4\Pi_g$      &$\rightarrow$  &$1\,^4\Delta_u$    \\  
   $1\,^2\Delta_u$   &$\rightarrow$  &$f\,^2\Pi_g$       \\
      $2\,^4\Pi_g$      &$\rightarrow$  &$B\,^4\Sigma_u^-$ \\ 
    \enddata
\end{deluxetable}
Since the dominant configurations change as $\cation$ dissociates---as highlighted 
in the early MRDCI work of \citet{Petrongolo81}, the valence-CI calculations of 
\citet{ballance01} and in more recent elaborate MRCI+Q 
calculations of \citet{Shi13}---only the initial and 
final molecular states are listed.

\begin{figure}[htp]
\plotone{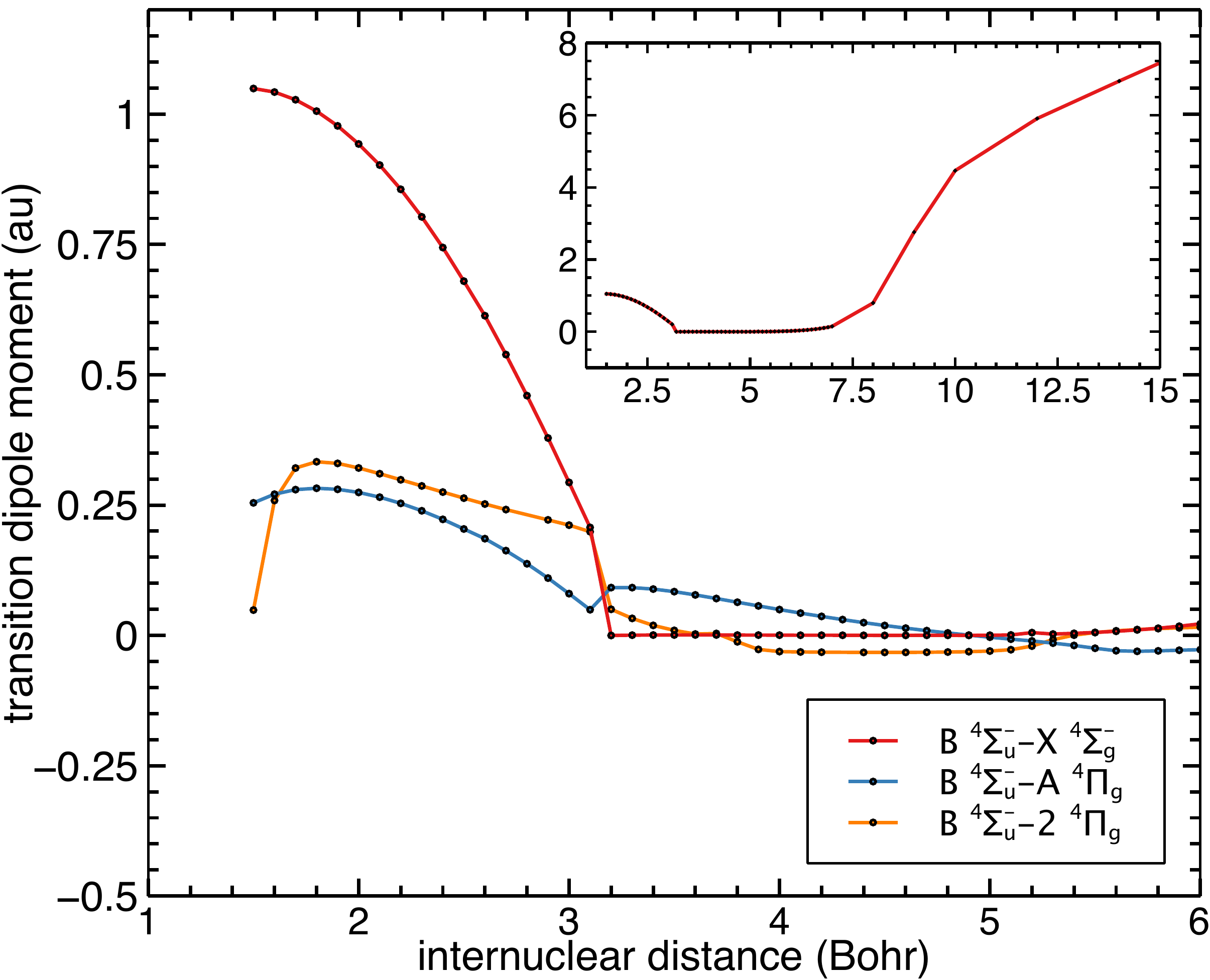}
\caption{Transition dipole moments for quartet transitions
involving the $B\,^4\Sigma_u^-$ state 
as functions of internuclear distance $R$:
$B\,^4\Sigma_u^-$-$X\,^4\Sigma_g^-$,
$B\,^4\Sigma_u^-$-$A\,^4\Pi_g$, and
$B\,^4\Sigma_u^-$-$2\,^4\Pi_g$.
The inset shows that the $B\,^4\Sigma_u^-$-$X\,^4\Sigma_g^-$ function
approaches $\textstyle \frac{1}{2}R$  for large $R$.
\label{fig:tdm-quartet-major}}
\end{figure}

\begin{figure}[htp]
\plotone{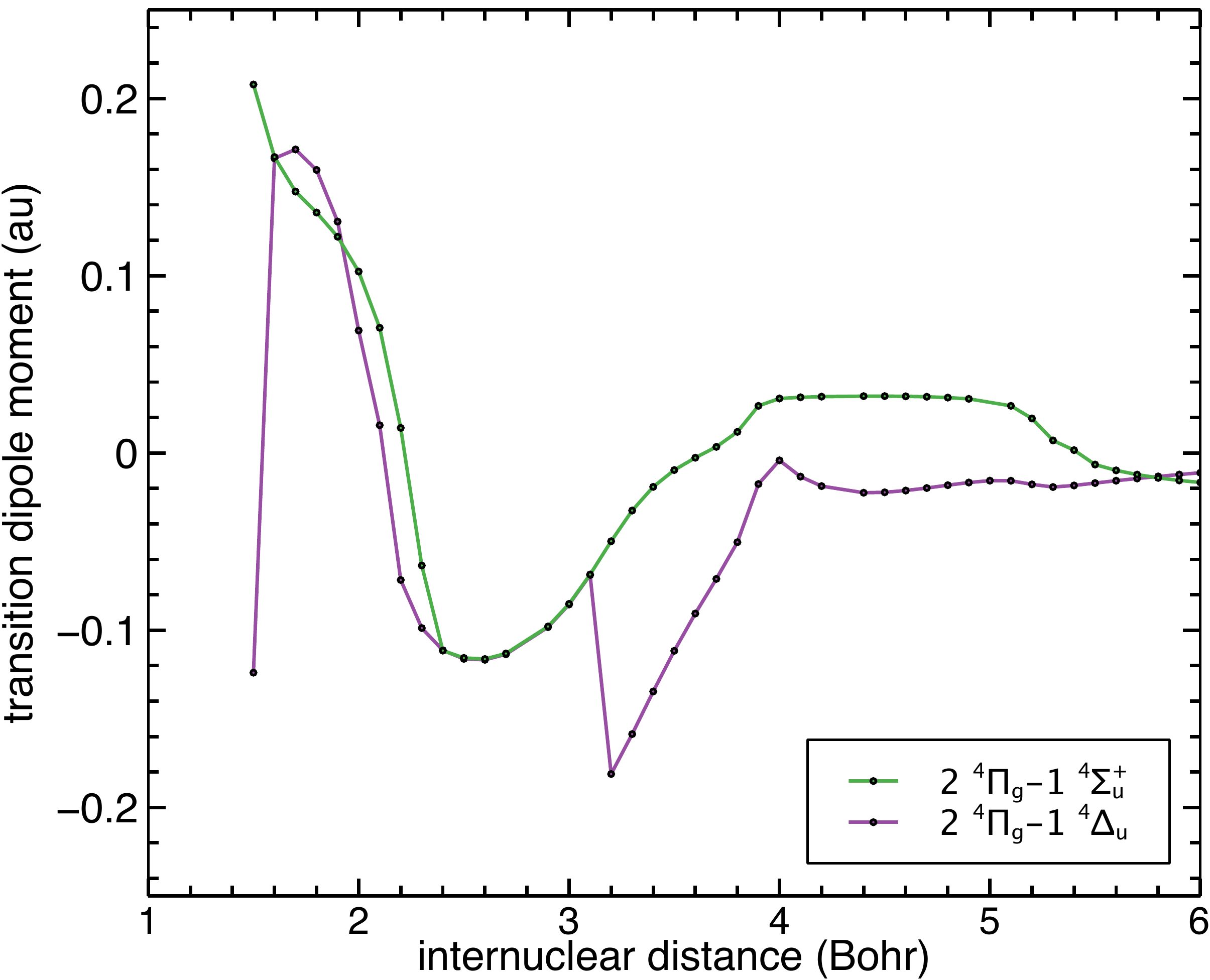}
\caption{Transition dipole moments for quartet transitions 
as a functions of internuclear distance $R$:
$2\,^4\Pi_g$-$1\,^4\Sigma_u^+$,
and
$2\,^4\Pi_g$-$1\,^4\Delta_u$.
\label{fig:tdm-quartet-minor}}
\end{figure}

\begin{figure}[htp]
\plotone{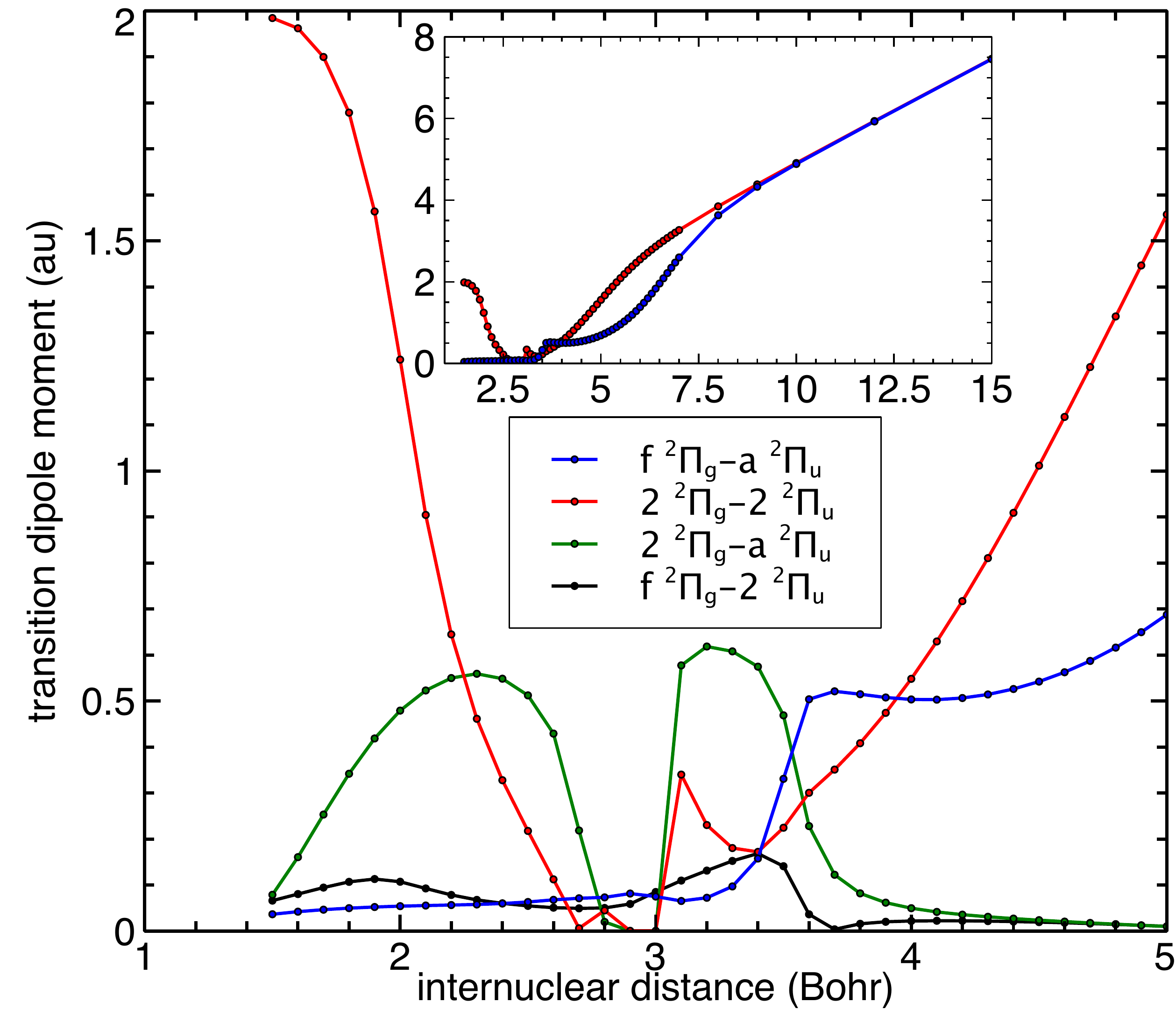}
\caption{Transition dipole moments between $^2\Pi_g$ and $^2\Pi_u$ states
as a function of internuclear distance $R$:
$f\,^2\Pi_g$-$a\,^2\Pi_u$,
$2\,^2\Pi_g$-2$^2\Pi_u$,
$2\,^2\Pi_g$-$a\,^2\Pi_u$, and
$f\,^2\Pi_g$-2$^2\Pi_u$.
The inset shows that the $2\,^2\Pi_g$-$a\,^2\Pi_u$
and $f\,^2\Pi_g$-$a\,^2\Pi_u$ functions
approach $\textstyle \frac{1}{2}R$  for large $R$.
\label{fig:tdm-doublet1}}
\end{figure}

\begin{figure}[htp]
\plotone{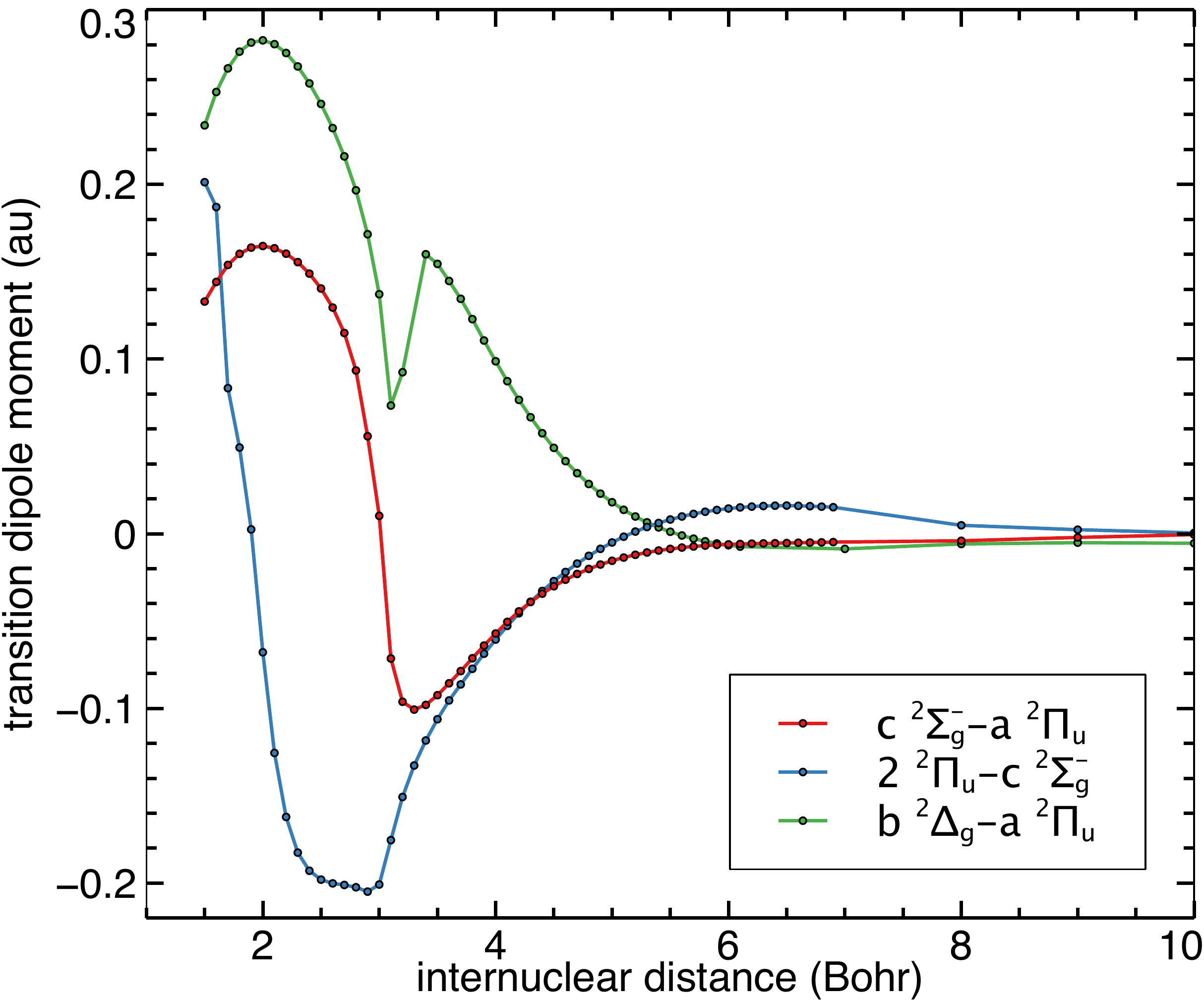}
\caption{Transition dipole moments as a function of internuclear distance $R$ between
$^2\Pi_u$ states and the $c\,^2\Sigma_g^-$ state and
for the $b^2\Delta_g$-$a\,^2\Pi_u$ transition.
\label{fig:tdm-doublet2}}
\end{figure}

\begin{figure}[htp]
\plotone{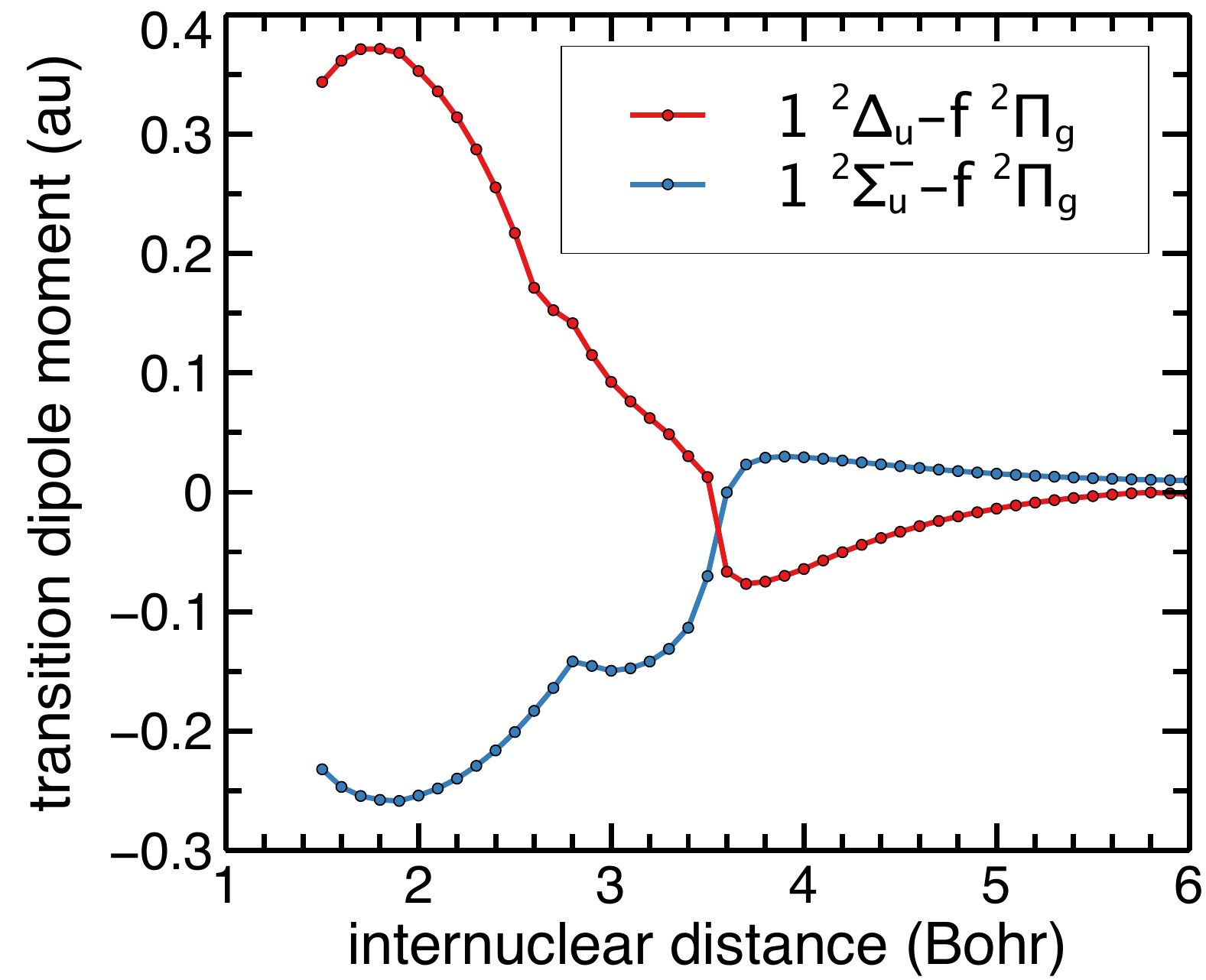}
\caption{Transition dipole moments for $f\,^2\Pi_g$-$1\,^2\Delta_u$
and
$f\,^2\Pi_g$-$1\,^2\Sigma_u^-$   
as a function of internuclear distance $R$.
\label{fig:tdm-doublet3}}
\end{figure}

For the quartet electronic states,
we selected the transitions
$B\,^4\Sigma_u^-$-$X\,^4\Sigma_g^-$, $B\,^4\Sigma_u^-$-$A\,^4\Pi_g$,
and $B\,^4\Sigma_u^-$-$2\,^4\Pi_g$, 
with our calculated TDMs shown in Fig.~\ref{fig:tdm-quartet-major},
and the
$2\,^4\Pi_g$-$1\,^4\Sigma_u^+$,
$2\,^4\Pi_g$-$1\,^4\Delta_u$,
$1\,^4\Sigma_u^+$-$A\,^4\Pi_g$
and
$1\,^4\Delta_u$-$A\,^4\Pi_g$
transitions, with our calculated TDMs
shown in Fig.~\ref{fig:tdm-quartet-minor}.
The $B\,^4\Sigma_u^-$-$X\,^4\Sigma_g^-$ TDM agrees
with the calculations of~\citet{Rosmus86},
which was given over the range $2<R<3.2\;a_0$.
For the doublet electronic states we selected the 
$f\,^2\Pi_g$-$a\,^2\Pi_u$, 
2$^2\Pi_g$-2$^2\Pi_u$,
2$^2\Pi_g$-$a\,^2\Pi_u$,
and
$f\,^2\Pi_g$-$2\,^2\Pi_u$ transitions,
where our calculated TDMs are shown in Fig.~\ref{fig:tdm-doublet1},
the 
$c\,^2\Sigma_g^-$-$a\,^2\Pi_u$-,
$2\,^2\Pi_u$-$c\,^2\Sigma_g^-$, and
$b\,^2\Delta_g$-$a\,^2\Pi_u$ transitions,
where our calculated TDMs are shown in Fig.~\ref{fig:tdm-doublet2},
and the
$f\,^2\Pi_g$-1$^2\Delta_g$
and
$f\,^2\Pi_g$-1$^2\Sigma_g^-$ transitions,
where our calculated TDMs are shown in Fig.~\ref{fig:tdm-doublet3}.
Our calculated value of the $f\,^2\Pi_g$-$a\,^2\Pi_u$
TDM at $R=2.4$ is 0.06, in atomic units, which agrees with the
estimate calculated 
at this single value of $R$
by \citet{Rosmus86}.

In general, the quartet and doublet TDMs are well-behaved at the lower
and upper limits of the ranges of $R$ for which they are calculated.
However, we note that for the quartet TDMs, Figs.~\ref{fig:tdm-quartet-major} and \ref{fig:tdm-quartet-minor},
there are sharp jumps around $R=3.2$.
We attribute this to changes in the dominant configurations
in the electronic wave functions around the four-way intersection of
the $A\,^4\Pi_g$, $B\,^4\Sigma_u^-$, $1\,^4\Delta_u$,
and $1^4\Sigma_u^+$ PECs, see Fig.~\ref{fig:pots-quartet}.
Moreover, the $1^4\Sigma_u^+$ and $1\,^4\Delta_u$ states are nearly degenerate~\citep{Petrongolo81,Shi13}
and the TDMs for 
$2\,^4\Pi_g$-$1\,^4\Sigma_u^+$ and
$2\,^4\Pi_g$-$1\,^4\Delta_u$
become equal over the range $2.4<R<3.1$.
In the case of the doublet TDMs, rapid variation around $R=2.9$ in the functions
involving the $a\,^2\Pi_u$ and $2\,^2\Pi_u$ states, Figs.~\ref{fig:tdm-doublet1} and \ref{fig:tdm-doublet2}, reflects
the avoided crossing between the PECs
where the dominant configuration of the $a\,^2\Pi_u$ state changes~\citep{Petrongolo81,ballance01}.
A more detailed analysis of the electronic
state configurations and their effect
on the TDMs should be carried out for spectroscopic applications,
such as for calculations of band oscillator strengths.
Such an analysis is unnecessary for our purposes,
because, as we will show, the most important factors leading to significant
radiative association cross sections are the lack of a barrier in the PEC
and a significant TDM at large $R$.
The $f\,^2\Pi_g$ entrance channel lacks a barrier and 
the TDMs for the $B\,^4\Sigma_u^-$-$X\,^4\Sigma_g^-$, 
$f\,^2\Pi_g$-$a\,^2\Pi_u$, and $2^2\Pi_g$-$2^2\Pi_u$ transitions approach $R/2$ for large $R$, 
as shown in the insets of Figs.~\ref{fig:tdm-quartet-major} and \ref{fig:tdm-doublet1},
similarly
to the $A$-$X$ TDM of $\textrm{O}_2^+$~\citep{Wetmore84}. 

The TDMs were fit to the form $1/R^4$, for $R \geq 20$,
except for the $B^4\Sigma_u^-$-$X^4\Sigma_g^-$, $f^2\Pi_g$-$a^2\Pi_u$,
and $2^2\Pi_g$-$2^2\Pi_u$ transitions 
which were fit to
$R/2$ at large nuclear separations.

\section{CROSS SECTIONS}\label{sec:cross-sections}

The radiative association cross sections for each transition
listed in Table~\ref{table:transitions}
contributing to process (\ref{RA}) were calculated for $^{12}\mathrm{C}$ nuclei 
as a function of the collision energy using the quantum-mechanical formalism,
which was previously applied to like-atom-ion radiative association for species without nuclear spin;
for $\mbox{He}$ and $\mbox{He}^+$~\citep{stancil93,augustovicova13} 
and for $\mbox{O}$ and $\mbox{O}^+$~\citep{BabbFan94}.

From the 24 molecular electronic states formed from $\Catom$ and $\Cion$
there are 72 possible approach channels 
considering that
due to the absence of nuclear spin  only half of the possible lambda doubling levels are populated for $\Lambda=1$ or $2$.
Thus the weight factor appearing in the cross section
for an entrance collision channel labeled by $i$ is $P_i=\textstyle \frac{1}{72}(2S_i+1)$.
Because the collisional cross sections will be mainly
due to partial waves with $N\gg 1$, we approximate $J=N$
and use 
Hund's
case~(b) coupling, where 
$N$ is the rotational
quantum number, $J$ is
the total angular momentum $|\mathbf{N}+\mathbf{S}|$, and
$S$ is the total electronic spin, 
as in our previous work on radiative association of $\mbox{O}$ and $\mbox{O}$~\citep{Babb1995}.
The reduced mass $\mu$ is $10937.35$ in units of the electron mass. 
As discussed in Sec.~\ref{sec:struct}, the potential energy curves and
transition dipole moment functions utilized were fit to appropriate long-range forms.
The cross sections were evaluated using the methods detailed in~\citet{Babb2019}.
For  collisions in the  $B\;^4\Sigma_u^-$ entrance channel only odd partial waves were treated,
while for the $1\;^4\Sigma_u^+$ and $c\,^2\Sigma_g^-$ entrance channels only even partial waves were treated.

\begin{figure}[htp]
\includegraphics[angle=000,width=6.5in]{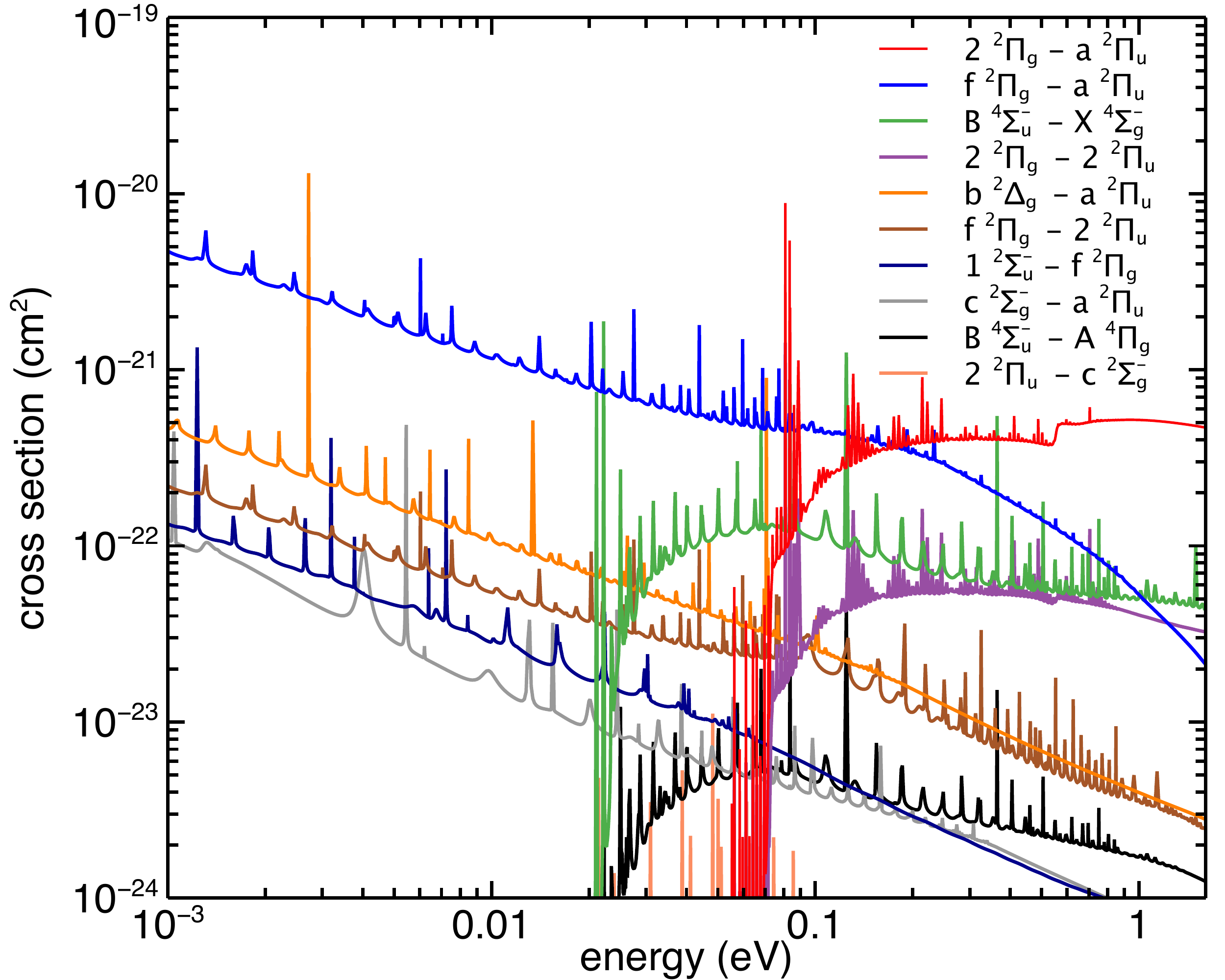}
\caption{Cross sections (in units of cm$^2$) as functions 
         of the collisional energy (in units of eV)  for  
         doublet and quartet transitions.
         The weight factors are set to unity in this plot.
\label{fig:cross}}
\end{figure}

The calculated cross sections are shown in Fig.~\ref{fig:cross}.
We do not plot the cross sections for the 
$2\,{}^2\Pi_u$-$c\,{}^2\Sigma_g^-$,
$2\,{}^4\Pi_g$-$1\,{}^4\Sigma_u^+$,
$2\,{}^4\Pi_g$-$1\,{}^4\Delta_u$, 
$1\,{}^2\Delta_u$-$f\,{}^2\Pi_g$,
and 
$2\,{}^4\Pi_g$-$B\,{}^4\Sigma_u^-$,
transitions, which were found to be insignificant.
Note that we also explored the $1\,{}^4\Sigma_u^+$-$A\,{}^4\Pi_g$
and $1\,^4\Delta_u$-$A\,^4\Pi_u$ transitions and
found that the TDMs are comparable in magnitude to those
shown in Fig.~\ref{fig:tdm-quartet-minor}. Therefore because
the transition energies are comparable to the states shown 
in that figure, we expect the cross sections to be insignificant.
\clearpage
\section{RATE COEFFICIENTS}
\begin{figure}[htp]
\includegraphics[angle=000,width=6.5in]{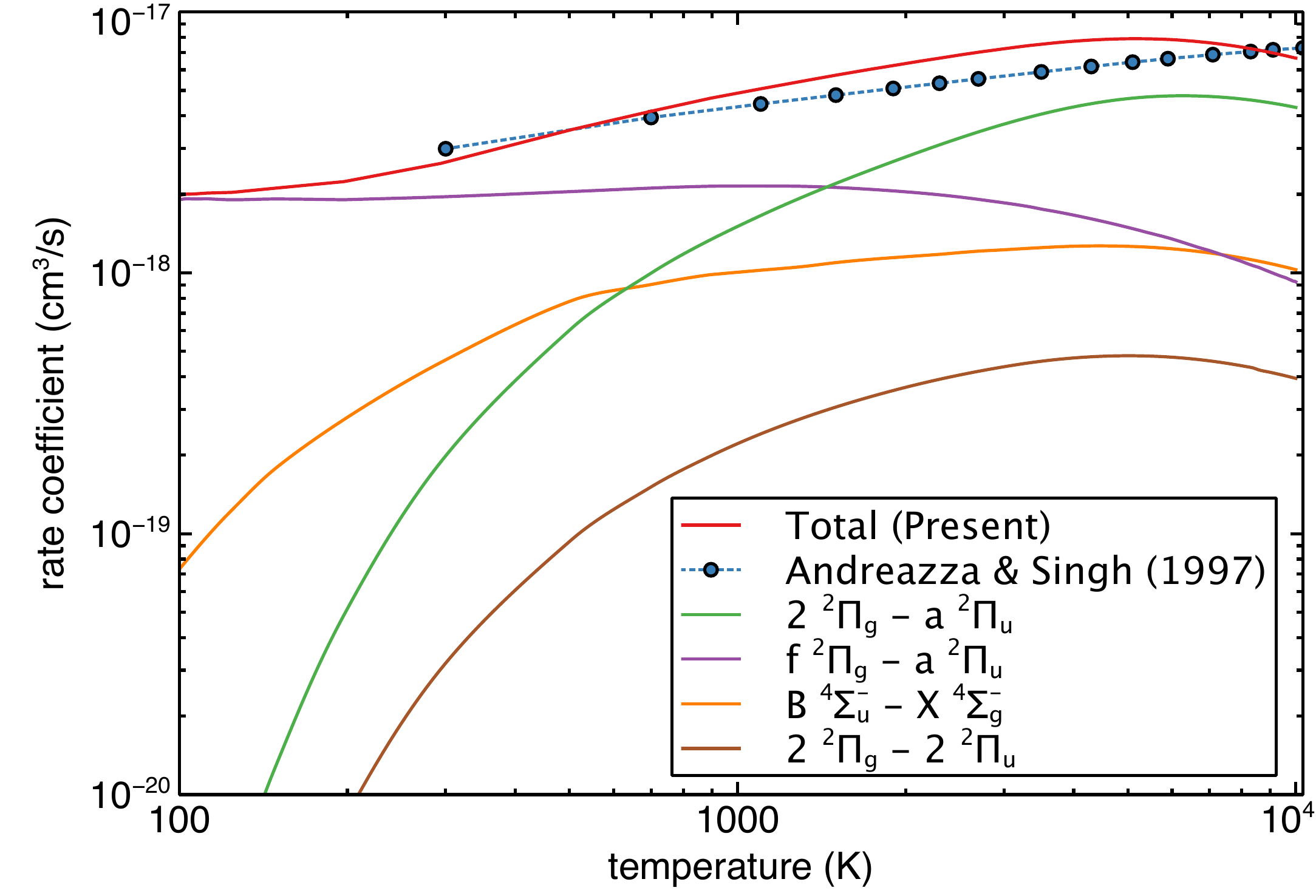}
\caption{
Rate coefficients (in units of $\rateunit$) as 
as functions of temperature (in units of K)  for 
the dominant channels contributing to the
radiative association association process (\ref{RA}).
The rate coefficients
for the four transitions and 
the sum are plotted in comparison to
the calculations of \citet{andreazza97}.
\label{fig:rates}}
\end{figure}

The rate coefficients were calculated by averaging the cross sections
over a Maxwellian velocity distribution.
The dominant entrance channels were found to be the doublet
$f\,^2\Pi_g$ and $2\,^2\Pi_g$ states and the quartet $B\,^4\Sigma_u^-$ state.
The dominant transitions are
$f\,^2\Pi_g$-$a\,^2\Pi_u$,
$2\,^2\Pi_g$-$2\,^2\Pi_u$,
$2\,^2\Pi_g$-$a\,^2\Pi_u$,
and $B\,^4\Sigma_u^-$-$X\,^4\Sigma_g^-$.
The rate coefficients
for these four transitions and
the total rate coefficients are shown in Fig.~\ref{fig:rates}
with the  rate coefficients 
for
the $B^4\Sigma_u^-$-$X^4\Sigma_g^-$ transition
from  \citet{andreazza97}.
The agreement between our total rate coefficients
and those from \citet{andreazza97} is good
but appears to be accidental.
\citet{andreazza97} selected the $B^4\Sigma_u^-$-$X^4\Sigma_g^-$ transition  based on a careful
analysis of available transition dipole moment data from the literature.
In particular, they determined that doublet transitions would not
contribute significantly compared to their predictions
for the strengths of the $B^4\Sigma_u^-$-$X^4\Sigma_g^-$ transition.
\citet{andreazza97} considered  the $f\,^2\Pi_g$-$a\,^2\Pi_u$ transition
but estimated that it was insignificant 
based on the relatively small TDM value at $R=2.4$ calculated
by \citet{Rosmus86}.
As discussed in Sec.~\ref{sec:struct},
our calculated TDM is in agreement
with the estimate given by \citet{Rosmus86}; however,
as illustrated in Fig.~\ref{fig:tdm-doublet1} we find that
the TDM of the $f\,^2\Pi_g$-$a\,^2\Pi_u$ transition grows linearly
with internuclear distance and
consequently  this is the most significant contribution
to process (\ref{RA})
at the lowest
energies because there is no barrier
in the  $f\,^2\Pi_g$ approach channel.
In contrast, while the $2\,^2\Pi_g$
and $B\,^4\Sigma_u^-$ approach channels
have barriers and won't contribute at the lowest energies,
for sufficient collisional 
energies the $2\,^2\Pi_g$-$2\,^2\Pi_u$ 
and $B\,^4\Sigma_u^-$-$X\,^4\Sigma_g^-$ transitions
become dominant.

The rate coefficients for the $B\,^4\Sigma_u^-$-$X\,^4\Sigma_g^-$  transition 
calculated by \citet{andreazza97}
increase slowly over the temperature
range from  300 to 14,700~K, varying from about $3\times10^{-18}\;\rateunit$ at 300~K
to $7.6\times 10^{-18}\;\rateunit$ at 14,700~K.
In the present calculations, there is a barrier of about 0.023~eV (effective
kinetic temperature of 270~K) at about $R=9\,a_0$,
in the $B\,^4\Sigma_u^-$ state.
Thus, as is evident in Fig.~\ref{fig:cross}, the
cross sections for the $B\,^4\Sigma_u^-$-$X\,^4\Sigma_g^-$
and $B\,^4\Sigma_u^-$-$A\,^4\Pi_g$ transitions rapidly
diminish for collisional energies
below 0.023~eV and these transitions will be inefficient
at temperatures below about 300~K, in marked contrast
to the calculations of \citet{andreazza97}, which do not show
a rapid decrease of the $B\,^4\Sigma_u^-$-$X\,^4\Sigma_g^-$ transition
rate coefficient  below 1000~K.
Therefore, any apparent agreement between
the total rate coefficient
obtained here and the calculations of \citet{andreazza97}, certainly
for temperatures below 1000~K, must be accidental.
Because of 
the similarity
between the 
$2\,^2\Pi_g$-$a\,^2\Pi_u$ 
and $B\,^4\Sigma_u^-$-$X\,^4\Sigma_g^-$
TDMs
and because the well depths of the $a\,^2\Pi_u$
and the $X\,^4\Sigma_g^-$ PECs are comparable
the earlier calculations of \citet{andreazza97} 
are in good agreement with
the present results, though the identifications
of the dominant formation channels for
the formation of $\cation$ via process (\ref{RA}) differ.

The total calculated rate coefficient $\alpha(T)$ is fit to better
than 11\% 
by the expression
\begin{equation}
\label{Present-fit}
    \alpha(T) = a + bT +cT^2 +dT^3\;\; (\rateunit), \qquad 100<T<10,000~K,
\end{equation}
with $a=1.9038\times 10^{-18}$,
$b=2.9628\times 10^{-21}$,
$c=-4.4649\times 10^{-25}$,
and
$d=1.9827\times 10^{-29}$.

\section{DISCUSSION}
We find that the rate coefficient for process (\ref{RA}) 
is about $2\times 10^{-18}\;\rateunit$ at 100~K.
It is important to note that the only prior existing calculation~\citep{andreazza97} 
is valid for temperatures greater than 300~K.
As we showed, the calculation of \citet{andreazza97} agrees fortuitously
with our calculation over the range $300<T<10000$~K and
they provided the fit
\begin{equation}
\label{ASfit}
\alpha_{\mathrm{AS97}}(T)= 4.01\times 10^{-18} (T/300)^{0.17} \exp(-101.52/T)\;\rateunit .
\end{equation}
In the context of astrochemical models, the fit provided by \citet{andreazza97},
which was limited to $300<T<41,000$~K, is implemented in
astrochemical databases that are applied to temperatures below 300~K.
Moreover,  applying the fit function from \citet{andreazza97},
for example, (incorrectly) to 100~K yields a rate coefficient 
that is a factor of two lower with an exponential decrease for $T<100$~K.
The fit, Eq.~(\ref{ASfit}), is listed as applicable for $10<T<41000$~K in the 
UMIST RATE12 file\footnote{\texttt{http://udfa.ajmarkwick.net/index.php?species=46}, accessed August 4, 2019.}, and was used, for example, in the photon-dominated
region code comparison study\footnote{Reaction rate file  \texttt{rate99\_edited\_incl\_crp.dat}  available at 
\texttt{https://zeus.ph1.uni-koeln.de/\-site/\-pdr-comparison/benchmark.htm},
accessed August 4, 2019.} with the same temperature validity listed~\citep{Rollig07}.
The present rate coefficients are fit, to within $4\%$
by the function
\begin{equation}
\label{Present-fit2}
\alpha_{\mathrm{low}}(T)\approx 2.55\times 10^{-18} (T/300)^{0.26}\;\rateunit\quad 100<T<300~\textrm{K}.
\end{equation}
Based on the trend of the $f\,^2\Pi_g$-$a\,^2\Pi_u$ cross section
with decreasing collision energy, see Fig.~\ref{fig:cross}, 
and estimating the rate coefficient
using $\alpha\sim \langle \sigma(E) v \rangle$, where $v=2E/\mu$ with $\mu$
the reduced mass, we can expect that the rate coefficient will be no greater
than $2\times 10^{-18}\;\rateunit$ at 10~K, though a detailed calculation
considering the fine structure of $\Catom$ and $\Cion$ should be performed
(the fine structure splitting of $\Cion$ is about $63.4\;\invcm$ or $\sim 91$~K).
The calculated rate coefficients for (\ref{RA}) are small, and the process will certainly
be unimportant when hydrogen is present.
Nevertheless, we conclude that the present rate coefficient will lead to an enhancement in $\cation$ production,
generally speaking, in astrochemical models, compared
to using the fit (\ref{ASfit}),
which is invalid for $T<300$~K.

\section{CONCLUSIONS}
We calculated the rate coefficients for the formation
of $\cation$ by radiative association of $\Catom$ and $\Cion$; process (\ref{RA}).
We substantially added to the available molecular
data on $\cation$ by calculating a large set of PECs
and TDMs.
We found that the dominant approach channels contributing
to process (\ref{RA}) are the doublet $f\,^2\Pi_g$ and $2\,^2\Pi_g$
states and that the quartet $B\,^4\Sigma_u^-$ approach channel
is of lesser importance.
The rate coefficient for process (\ref{RA}) should not decrease exponentially
as temperature decreases because there is no barrier in the dominant $f\,^2\Pi_g$ 
entrance channel.
As a consequence, the present rate coefficients will lead
to enhanced production of $\cation$ in astrophysical models,
though the net impact may be insignificant when hydrogen is present.

\acknowledgments

We thank Dr. Romane Le Gal for helpful discussions.
This work was supported by a Smithsonian Scholarly Studies grant.
ITAMP is supported in part by NSF Grant No.\ PHY-1521560.
BMMcL acknowledges support by
the ITAMP visitor's program, by the University of 
Georgia at Athens for the award of an adjunct professorship, 
and by Queen's University Belfast for a visiting research fellowship (VRF). 
We thank Captain Thomas J. Lavery, USN, Ret., for his constructive comments that
enhanced the quality of this manuscript.
The authors acknowledge this research 
used grants of computing time at the National
Energy Research Scientific Computing Centre (NERSC), which is supported
by the Office of Science of the U.S. Department of Energy
(DOE) under Contract No. DE-AC02-05CH11231.
The authors gratefully acknowledge the Gauss Centre for 
Supercomputing e.V. (www.gauss-centre.eu) 
for funding this project by providing computing time on the GCS Supercomputer 
HAZEL HEN at H\"{o}chstleistungsrechenzentrum Stuttgart (www.hlrs.de).
\software{  
          \textsc{molpro} (v.~2015.1) \citep{molpro2015}
          }

\clearpage

\begin{thebibliography}{}
\expandafter\ifx\csname natexlab\endcsname\relax\def\natexlab#1{#1}\fi
\providecommand{\url}[1]{\href{#1}{#1}}
\providecommand{\dodoi}[1]{doi:~\href{http://doi.org/#1}{\nolinkurl{#1}}}
\providecommand{\doeprint}[1]{\href{http://ascl.net/#1}{\nolinkurl{http://ascl.net/#1}}}
\providecommand{\doarXiv}[1]{\href{https://arxiv.org/abs/#1}{\nolinkurl{https://arxiv.org/abs/#1}}}

\bibitem[{Andreazza \& Singh(1997)}]{andreazza97}
Andreazza, C.~M., \& Singh, P.~D. 1997, MNRAS, 287, 287,
  \dodoi{10.1093/mnras/287.2.287}

\bibitem[{Augustovi\v{c}ov\'a {et~al.}(2013)Augustovi\v{c}ov\'a, \v{S}pirko,
  Kraemer, \& Sold\'an}]{augustovicova13}
Augustovi\v{c}ov\'a, L., \v{S}pirko, V., Kraemer, W.~P., \& Sold\'an, P. 2013,
  A\&A, 553, A42, \dodoi{10.1051/0004-6361/201220957}

\bibitem[{Babb {et~al.}(1994)Babb, Fan, \& Dalgarno}]{BabbFan94}
Babb, J., Fan, Z., \& Dalgarno, A. 1994, J. Quant. Spectrosc. Rad. Trans., 52,
  161, \dodoi{10.1016/0022-4073(94)90006-X}

\bibitem[{Babb \& Dalgarno(1995)}]{Babb1995}
Babb, J.~F., \& Dalgarno, A. 1995, \pra, 51, 3021,
  \dodoi{10.1103/PhysRevA.51.3021}

\bibitem[{Babb {et~al.}(2019)Babb, Smyth, \& McLaughlin}]{Babb2019}
Babb, J.~F., Smyth, R.~T., \& McLaughlin, B.~M. 2019, \apj, 876, 38,
  \dodoi{10.3847/1538-4357/ab1088}

\bibitem[{Ballance \& McLaughlin(2001)}]{ballance01}
Ballance, C.~P., \& McLaughlin, B.~M. 2001, J. Phys. B: At. Molec. Opt. Phys.,
  34, 1201, \dodoi{10.1088/0953-4075/34/7/304}

\bibitem[{{Boudjarane} {et~al.}(1995){Boudjarane}, {Carr\'e}, \&
  {Larzilli\`ere}}]{boudjarane95}
{Boudjarane}, K., {Carr\'e}, M., \& {Larzilli\`ere}, M. 1995, Chem. Phys.
  Lett., 243, 571, \dodoi{10.1016/0009-2614(95)00888-B}

\bibitem[{Bruna \& Wright(1992)}]{bruna92}
Bruna, P.~J., \& Wright, J.~S. 1992, \jcp, 96, 1630,
  \dodoi{10.1021/j100183a027}

\bibitem[{Burdakova {et~al.}(2019)Burdakova, Nyman, \& Stoecklin}]{burdakova19}
Burdakova, D., Nyman, G., \& Stoecklin, T. 2019, \mnras, 485, 5874,
  \dodoi{10.1093/mnras/stz795}

\bibitem[{Celii \& Maier(1990)}]{Celii1990}
Celii, F., \& Maier, J. 1990, Chem. Phys. Lett., 166, 517,
  \dodoi{10.1016/0009-2614(90)87144-G}

\bibitem[{{Chabot} {et~al.}(2013){Chabot}, {B{\'e}roff}, {Gratier}, {Jallat},
  \& {Wakelam}}]{Chabot13}
{Chabot}, M., {B{\'e}roff}, K., {Gratier}, P., {Jallat}, A., \& {Wakelam}, V.
  2013, \apj, 771, 90, \dodoi{10.1088/0004-637X/771/2/90}

\bibitem[{{Chiu}(1973)}]{Chiu73}
{Chiu}, Y.-N. 1973, \jcp, 58, 722, \dodoi{10.1063/1.1679259}

\bibitem[{Das \& Thakkar(1998)}]{DasTha98}
Das, A.~K., \& Thakkar, A.~J. 1998, J. Phys. B: At. Molec. Opt. Phys., 31,
  2215, \dodoi{10.1088/0953-4075/31/10/011}

\bibitem[{{Federman} \& {Huntress Jr.}(1989)}]{federman89}
{Federman}, S.~R., \& {Huntress Jr.}, W.~T. 1989, \apj, 338, 140,
  \dodoi{10.1086/167187}

\bibitem[{Freed {et~al.}(1982)Freed, Oka, \& Suzuki}]{Freed82}
Freed, K.~F., Oka, T., \& Suzuki, H. 1982, \apj, 263, 718,
  \dodoi{10.1086/160543}

\bibitem[{Gentry \& Giese(1977)}]{GenGie77}
Gentry, W.~R., \& Giese, C.~F. 1977, \jcp, 67, 2355, \dodoi{10.1063/1.435072}

\bibitem[{{Guzm{\'a}n} {et~al.}(2015){Guzm{\'a}n}, {Pety}, {Goicoechea},
  {Gerin}, {Roueff}, {Gratier}, \& {{\"O}berg}}]{Guzman15}
{Guzm{\'a}n}, V.~V., {Pety}, J., {Goicoechea}, J.~R., {et~al.} 2015, \apjl,
  800, L33, \dodoi{10.1088/2041-8205/800/2/L33}

\bibitem[{Helgaker {et~al.}(2000)Helgaker, J\o{}rgensen, \&
  Olsen}]{Helgaker2000}
Helgaker, T., J\o{}rgensen, P., \& Olsen, J. 2000, {Molecular
  Electronic-Structure Theory} (New York, USA: Wiley)

\bibitem[{Herzberg(1950)}]{herz50}
Herzberg, G. 1950, {Spectra of Diatomic Molecules} (New York, USA: Van
  Nostrand)

\bibitem[{Hirooka {et~al.}(2014)Hirooka, Sato, Ishihara, Yabuuchi, \&
  Tanaka}]{hirooka2014}
Hirooka, Y., Sato, H., Ishihara, K., Yabuuchi, T., \& Tanaka, K. 2014, Nucl.
  Fusion, 54, 022003, \dodoi{10.1088/0029-5515/54/2/022003}

\bibitem[{Maier \& R\"osslein(1988)}]{maier88}
Maier, J.~P., \& R\"osslein, M. 1988, \jcp, 88, 4614, \dodoi{10.1063/1.453774}

\bibitem[{Miller \& Kelly(1972)}]{MilKel72}
Miller, J.~H., \& Kelly, H.~P. 1972, \pra, 5, 516,
  \dodoi{10.1103/PhysRevA.5.516}

\bibitem[{Nyman {et~al.}(2015)Nyman, Gustafsson, \& Antipov}]{Nyman15}
Nyman, G., Gustafsson, M., \& Antipov, S.~V. 2015, Int. Rev. Phys. Chem., 34,
  385, \dodoi{10.1080/0144235X.2015.1072365}

\bibitem[{Petrongolo {et~al.}(1981)Petrongolo, Bruna, Peyerimhoff, \&
  Buenker}]{Petrongolo81}
Petrongolo, C., Bruna, P.~J., Peyerimhoff, S.~D., \& Buenker, R.~J. 1981, \jcp,
  74, 4594, \dodoi{10.1063/1.441648}

\bibitem[{Rampino {et~al.}(2016)Rampino, Pastore, Garcia, Pacifici, \&
  Lagan{\`a}}]{Rampino16}
Rampino, S., Pastore, M., Garcia, E., Pacifici, L., \& Lagan{\`a}, A. 2016,
  \mnras, 460, 2368, \dodoi{10.1093/mnras/stw1116}

\bibitem[{R{\"o}llig {et~al.}(2007)R{\"o}llig, Abel, Bell, Bensch, Black,
  Ferland, Jonkheid, Kamp, Kaufman, Le~Bourlot, Le~Petit, Meijerink, Morata,
  Ossenkopf, Roueff, Shaw, Spaans, Sternberg, Stutzki, Thi, van Dishoeck, van
  Hoof, Viti, \& Wolfire}]{Rollig07}
R{\"o}llig, M., Abel, N.~P., Bell, T., {et~al.} 2007, \aap, 467, 187,
  \dodoi{10.1051/0004-6361:20065918}

\bibitem[{Rosmus {et~al.}(1986)Rosmus, Werner, Reinsch, \& Larsson}]{Rosmus86}
Rosmus, P., Werner, H.-J., Reinsch, E.-A., \& Larsson, M. 1986, J. Electron.
  Spectrosc., 41, 289, \dodoi{10.1016/0368-2048(86)85009-5}

\bibitem[{Shi {et~al.}(2013)Shi, Niu, Sun, \& Zhu}]{Shi13}
Shi, D., Niu, X., Sun, J., \& Zhu, Z. 2013, J. Phys. Chem. A, 117, 2020,
  \dodoi{10.1021/jp312670v}

\bibitem[{{Solomon} \& {Klemperer}(1972)}]{Solomon72}
{Solomon}, P.~M., \& {Klemperer}, W. 1972, \apj, 178, 389,
  \dodoi{10.1086/151799}

\bibitem[{{Stancil} {et~al.}(1993){Stancil}, {Babb}, \& {Dalgarno}}]{stancil93}
{Stancil}, P.~C., {Babb}, J.~F., \& {Dalgarno}, A. 1993, \apj, 414, 672,
  \dodoi{10.1086/173113}

\bibitem[{Tarsitano {et~al.}(2004)Tarsitano, Neese, \& Oka}]{tarsitano04}
Tarsitano, C.~G., Neese, C.~F., \& Oka, T. 2004, \jcp, 121, 6290,
  \dodoi{10.1063/1.1787493}

\bibitem[{Werner {et~al.}(2012)Werner, Knowles, Knizia, Manby, \&
  Sch{\"u}tz}]{molpro2012}
Werner, H.-J., Knowles, P.~J., Knizia, G., Manby, F.~R., \& Sch{\"u}tz, M.
  2012, WIREs Comput. Mol. Sci., 2, 242, \dodoi{10.1002/wcms.82}

\bibitem[{Werner {et~al.}(2015)Werner, Knowles, Knizia, Manby, {Sch\"{u}tz},
  {et~al.}}]{molpro2015}
Werner, H.-J., Knowles, P.~J., Knizia, G., {et~al.} 2015, \textsc{molpro},
  version 2015.1, a package of \textit{ab initio} programs,  Cardiff, UK:
  http://www.molpro.net

\bibitem[{{Wetmore} {et~al.}(1984){Wetmore}, {Fox}, \& {Dalgarno}}]{Wetmore84}
{Wetmore}, R.~W., {Fox}, J.~L., \& {Dalgarno}, A. 1984, \planss, 32, 1111,
  \dodoi{10.1016/0032-0633(84)90136-3}

\end{thebibliography}
\end{document}